\documentclass[12pt,preprint]{aastex}
\usepackage{natbib}
\begin{document}

\title{A Very High Proper Motion Star and the First L dwarf in the {\it Kepler} Field}

\author{John E. Gizis, Nicholas W. Troup}
\affil{Department of Physics and Astronomy, University of Delaware, 
Newark, DE 19716}

\author{Adam J. Burgasser}
\affil{Center for Astrophysics and Space Science, University of California San Diego, La Jolla, CA 92093}
  
\begin{abstract}
We report two nearby high proper motion dwarfs of special interest identified using the Preliminary Data Release of the Wide-field Infrared Survey Explorer (WISE) and the Two Micron All-Sky Survey (2MASS).  WISEP J191239.91-361516.4 has a motion of 2.1 arcseconds per year. Photometry identifies it as a mid-M dwarf. WISEP J190648.47+401106.8 is a spectroscopically confirmed L1 dwarf in the {\it Kepler Mission} field with a motion of 0.48 arcseconds per year. The estimated distance is 17 parsecs. Both lie at relatively low galactic latitudes and demonstrate the possibility of discovering proper motion stars independently of the historic photographic sky surveys.
\end{abstract}


\section{Introduction}

The identification of high proper motion stars is an important tool to identify neighboring stars. The LHS Catalog \citep{lhscat} listed over four thousand stars with proper motions near or above half an arcsecond per year but only 73 stars with motions above two arcseconds per year. Digitized photographic sky surveys and the more recent digital optical and near-infrared sky surveys have allowed many new proper motion stars to be identified \citep{2002AJ....124.1190L,2005AJ....129.1483L,2005AJ....129..413S,2005AJ....130.1247L,2005A&A...435..363D,2007AJ....133.2898F,2008AJ....135.2177L,2009AJ....137..304S,2010ApJS..190..100K}. Currently, the Simbad database lists some 108 stars (including white dwarfs and brown dwarfs) with motions above two arcseconds per year.

Nevertheless, high proper motion stars remain to be identified. At low galactic latitudes, crowding makes it difficult to correctly pair stars, even though important progress has been made to bring the completeness in the Galactic Plane to at least 90\% \citep{2005AJ....129.1483L} and perhaps as high as 99\% \citep{2002AJ....124.1190L}. \citet{2005AJ....129.1483L} also estimate that incompleteness sets in at  $V=19$ for low galactic latitudes ($|b|<15$) but $V=20$ at high latitudes, and remark that independent surveys are needed to fully assess completeness of their catalog.  
Unfortunately, the vast number of reddened background stars makes it difficult to identify cool, nearby objects by photometry alone. A second potential source of incompleteness  is that for very large motions, the stars may move so far that automated pairing programs fail. The relatively recent discovery of such objects as a white dwarf moving at 2.55 "/yr \citep{2005ApJ...633L.121L} and an M6.5 dwarf moving at 5.05 "/yr \citep{2003ApJ...589L..51T}, both detectable on the photographic plate surveys, are illustrative.  Third, cooler objects, particularly brown dwarfs, are too faint for photographic plates, though these can often be selected by infrared colors alone.

These problems can be addressed with a new infrared sky survey. The Wide-field Infrared Survey Explorer (WISE) has surveyed the entire sky in four mid-infrared filters \citep{2010AJ....140.1868W}, allowing a comparison with the near-infrared Two Micron All-Survey (2MASS) \citep{2mass}, even in the Galactic Plane. The WISE Explanatory Supplement demonstrates that the matching of the two surveys is good to 0.2" for high signal-to-noise stars, and that 99.9\% of WISE stars match 2MASS sources to within 3 arcseconds. The time baseline is approximately a decade, so objects with $\mu \gtrsim 0.3$ "/yr are expected to move out of the matching window. The high reliability and completeness of WISE and 2MASS allow us to search for previously unidentified proper motion stars. In this {\it Letter}, we report on two particularly important sources. 

\section{Data Analysis}

Our strategy is based on the fact that nearby hydrogen-burning stars should appear as relatively bright, high signal-to-noise sources in both WISE and 2MASS.  We initially obtained a list of possible proper motion stars by querying the WISE Preliminary Source Catalog for stars with $5 < W_1 < 12$ that lacked 2MASS pairings within three arcseconds.  We then matched this list with the \citet{2005AJ....129.1483L} catalog of stars with motions greater than 0.15 arcseconds per year. We found that nearly all LSPM stars had $0.1 < W_1 - W_2 < 0.3$ and $-0.1 < W_1 - W_3 < 0.7$ (with the exception of a few very bright stars affected by saturation), and so we applied these color cuts to the sample. This selection for initial analysis, it should be noted, is intended to select main sequence stars and will likely exclude white dwarfs and brown dwarfs.  We furthermore excluded known stars by requiring no Simbad sources within 12 arcseconds and no matches to the PPMXL catalog \citep{2010AJ....139.2440R}. This left only 118 sources, each of which were examined in the Digitized Sky Survey, 2MASS and WISE images. Because we had made no additional cuts based on WISE source quality, this list included many bright star artifacts, as expected in the Preliminary processing. However, we also found we recovered a number of known proper motion stars, such as  the 5 "/yr star SO 025300.5+165258 \citep{2003ApJ...589L..51T} [WISEP J025303.27+165214.2] which move quickly enough that they were not excluded by the Simbad query, as well as previously unknown stars.  We extended this analysis to $|b| > 10$ without difficulty.  For yet lower galactic latitudes, $|b| < 10$, we found we could exclude most of the bright star artifacts, and none of the proper motion stars, by requiring that the source not be extended ($ext\_flg = 0$) in WISE. We also had to drop the W3 color cuts, because the 12 micron data do not reach as deep in regions of high background.  No matches at the lowest latitude to PPMXL were attempted given its potential unreliability in crowded fields. The resulting 492 sources were also examined by eye, and 229 appear to be genuine proper motion objects. The main sources of false matches in the Plane were apparently real but non-moving sources whose 2MASS detections were masked or flagged due to nearby bright stars in the crowded fields. As expected, most of the genuine high proper motion stars were listed by \citet{lhscat} or more recent publications \citep{2005AJ....130.1247L,2008AJ....135.2177L}.  Complete details and a full list of detected stars will be presented in a future publication, where we will assess completeness and reliability. 

Only one previously unknown source with motion above two arcseconds per year was detected in the 58\% of the sky in the Preliminary release --- that is, we are able identify pairings for all other bright WISE stars, as we define them above. WISEP J191239.91-361516.4  is at $b = -19.6$ and lies in a moderately crowded field.  We also discuss a second source, WISEP J190648.47+401106.8, with a proper motion of 0.48 "/yr at $b=14.4$. Further analysis of the other proper motion stars detected and the effect of relaxing the color constraint is ongoing and will be reported in a later paper.  

\section{Discussion}

\begin{figure}
%\epsscale{0.7}
\plotone{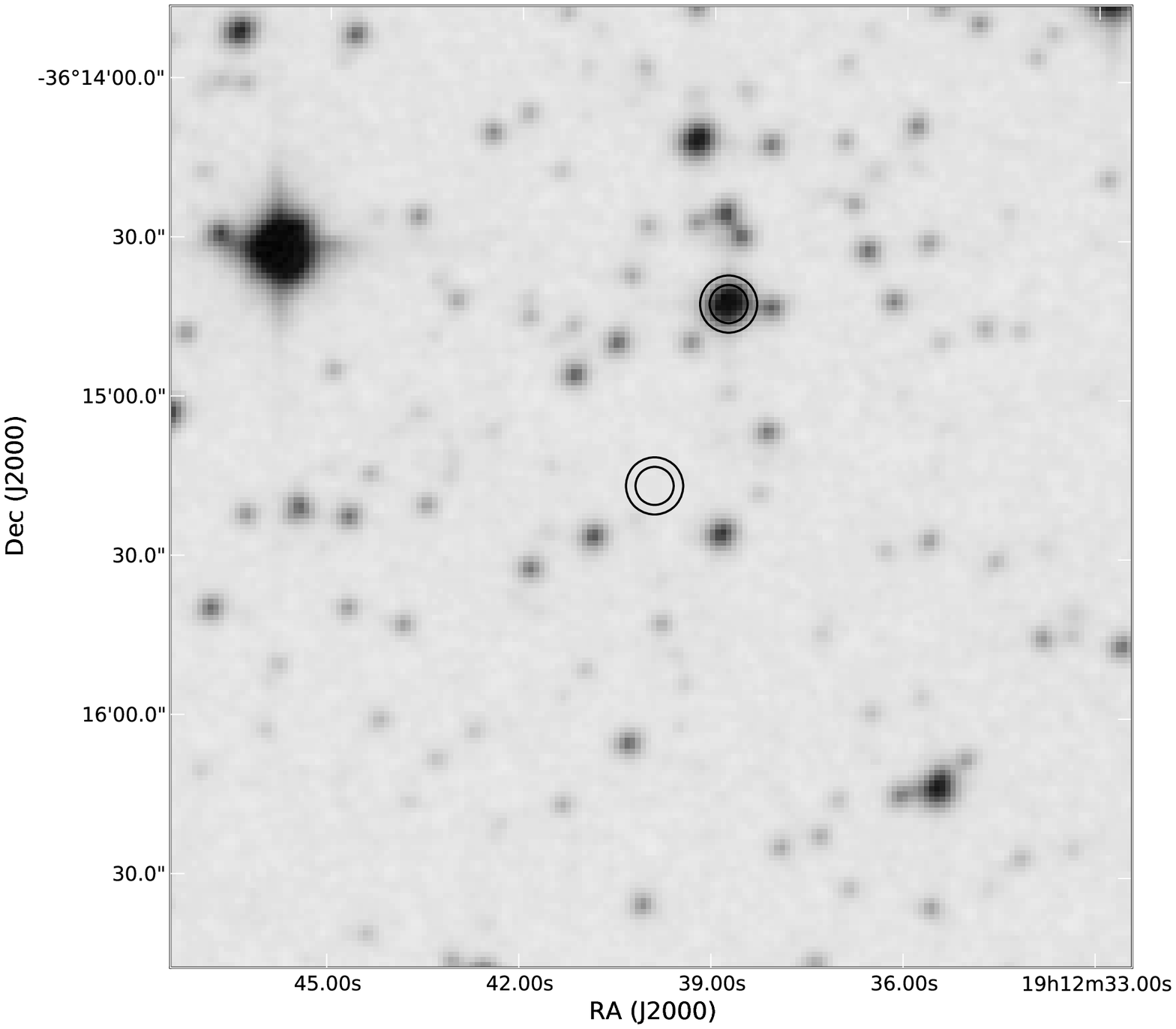}
\caption{Finder chart for WISEP J191239.91-361516.4. This image is the DSS scan of the 27 July 1992  AAO-SES red photographic plate. Both the 1992 position and the 2010 WISE position are marked with circles. 
\label{fig0}}
\end{figure}

\begin{figure}
%\epsscale{0.7}
\plotone{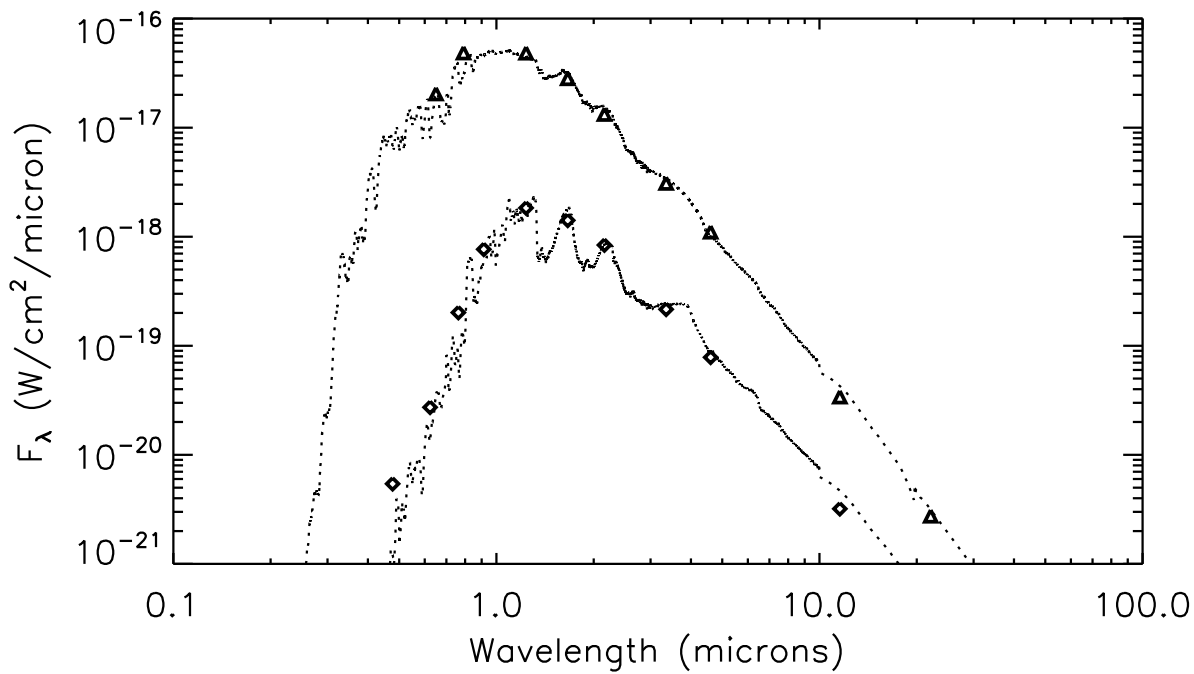}
\caption{The observed flux densities for the two stars (WISEP J191239.91-361516.4 as triangles, and WISEP  J190648.47+401106.8 as diamonds.) Magnitude zeropoints are from \citet{2009ApJS..182..543A} (SDSS), \citet{2000A&AS..141..313F} (DENIS), \citet{1992AJ....104.1650C} (2MASS), and \citet{2010AJ....140.1868W} (WISE). Model spectra for 3200K \citep{1999ApJ...512..377H} and 2000K \citep{dusty}, shown as dotted lines, are illustrative only.\label{fig1}}
\end{figure}

We identify WISEP J191239.91-361516.4 with 2MASS J19123922-3614555, DENIS 191239.2-361455, and USNO-B 0537-0751534. A finder chart is shown in Figure~\ref{fig0}. The WISE and 2MASS photometry and astrometry are listed in Table~1, and in addition, $I=10.98$ from DENIS \citep{1997Msngr..87...27E} and $R_F \approx 12.6$  from USNO-B \citep{2003AJ....125..984M}. The WISE observations are from 1 April 2010 to  7 April 2010 and the 2MASS observations are from 25 June 1999. Since the WISE positions are tied directly to the 2MASS positions, we can compute the proper motion of 2.09 "/yr directly.  The reported WISE-2MASS positional uncertainties of 0.2" lead to an uncertainty of the proper motions of 0.02 "/yr. (We can also directly compare WISE-2MASS proper motions to those reported in the LSPM catalog, and we find that the standard deviation of the differences is 0.013 "/yr in each coordinate, which can be attributed primarily to the WISE-2MASS uncertainties.) The photometry is consistent with a mid-M ($\sim$ M4) dwarf. The photometry and a 3200K model atmosphere \citep{1999ApJ...512..377H} are plotted in Figure~\ref{fig1}. Comparing to the parallax sample of \citet{2006AJ....132.1234C}, we expect $M_J \approx 9.0$ for $I-J=1.2$ for a disk main sequence dwarf. This suggests a distance of 13 parsecs and $v_{tan} \approx 120$ km/s, but if the star is metal-poor, it would be closer with a lower velocity, and a distance within 10 parsecs is possible. (If an equal-luminosity binary, it may be more distant.) A trigonometric parallax is needed.  According to Simbad, it is the 103rd fastest proper motion star known.

\begin{figure}
%\epsscale{0.7}
\plotone{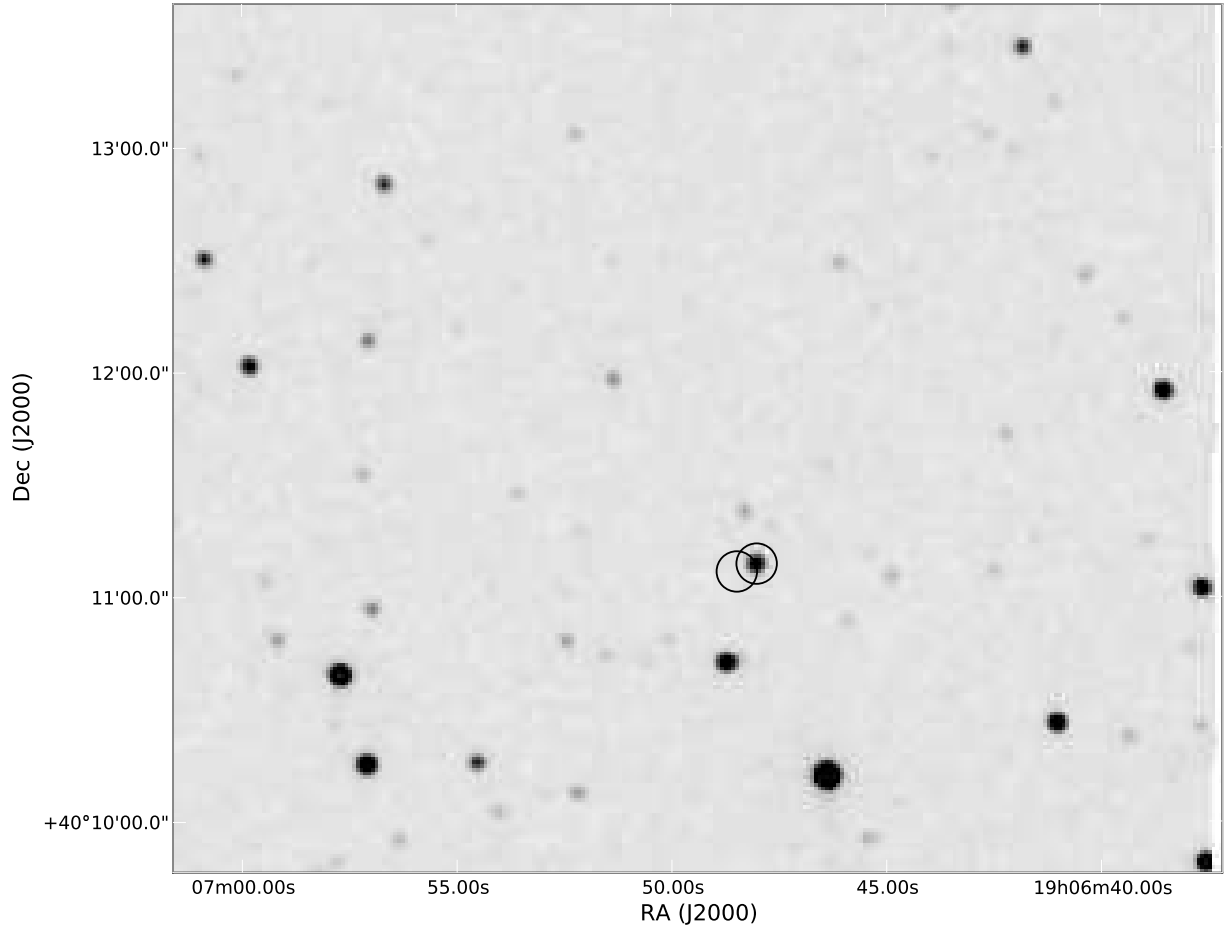}
\caption{Finder chart for WISEP J190648.47+401106.8. This is the 2MASS J-band image taken on 23 May 1998. Both the 1998 2MASS position and 2010 WISE position are marked by circles.
\label{fig3}}
\end{figure}

\begin{figure}
%\epsscale{0.7}
\plotone{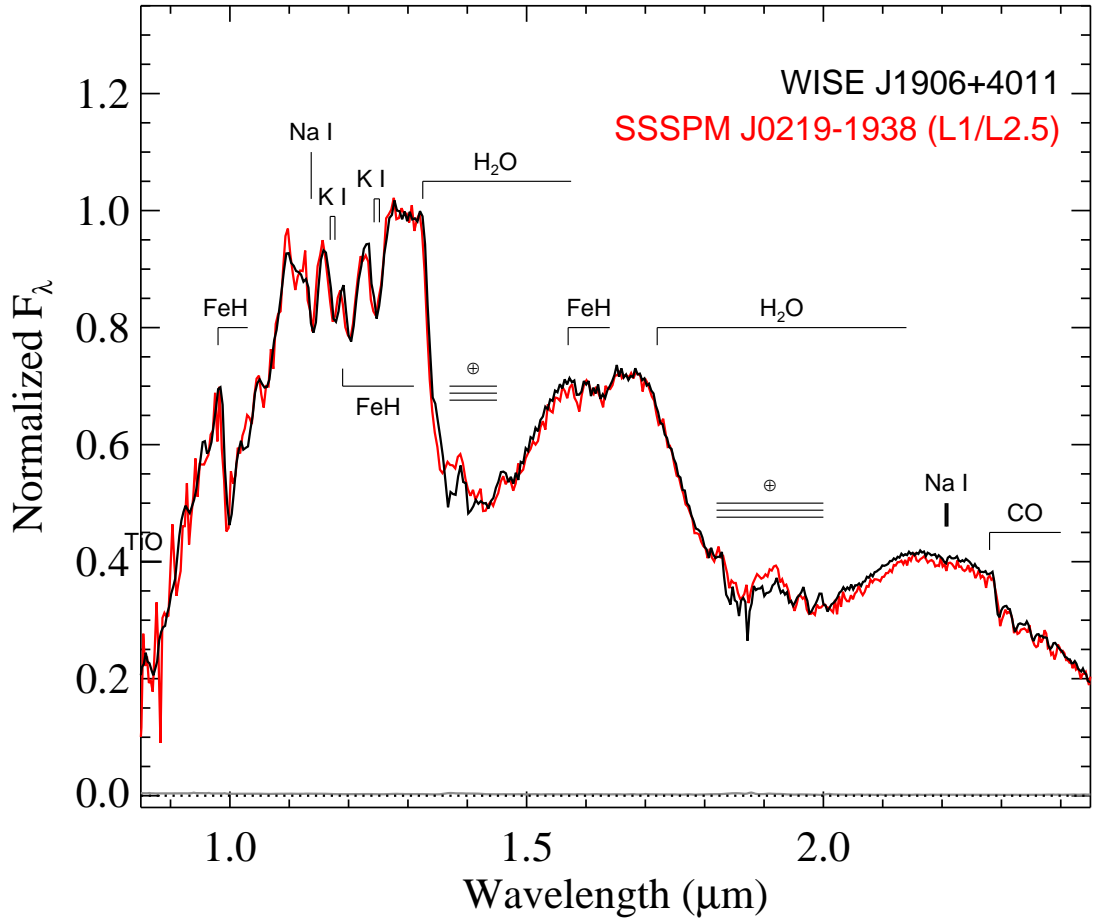}
\caption{The SpeX spectrum of WISEP  J190648.47+401106.8 compared to the L dwarf SSSPM J0219-1939, classified as L1 in the optical and L2.5 in the near-infrared \citet{ldwarf}.\label{fig2}}
\end{figure}

We identify WISEP J190648.47+401106.8 (hereafter W1906+40) with 2MASS J19064801+4011089 and SDSS J190648.29+401107.6 \citep{2009ApJS..182..543A}.  The ten photometric observations are shown in Figure~\ref{fig1}, with $g=22.357$, $r=20.029$, $i=17.419$, and $z=15.578$. The WISE observations are from 16 April 2010 to 22 April 2010 and the 2MASS observations are from 23 May 1998. A finder chart is shown in Figure~\ref{fig3}. Both $i-z = 1.84$ and $J-K_s= 1.31$ are consistent with a late-M or an early-L type \citep{2010AJ....139.1808S}. A low-resolution IRTF SpeX \citep{spex} spectrum was obtained on 19 April 2011 and processed using SpexTool \citep{vacca,spextool}.  It is shown in Figure~\ref{fig2}.  We classify W1906+40 as spectral type L1. The source may be a hydrogen-burning star or a brown dwarf. According to the relations of \citet{cruz}, the distance is $16.6\pm1.9$ parsecs and the tangential velocity is $\sim 38$ km/s, but again, a trigonometric parallax is needed.  The most compelling property of this nearby L dwarf is that it lies in the Kepler Mission field \citep{2010ApJ...713L..79K}, as Kepler ID 4996077, making it the coolest dwarf known in the field. It was not on the Kepler observing list, but examination of the Kepler Full Field Images (FFI) shows that W1906+40 is detected. DDT monitoring with Kepler will provide a unique time series for an ultracool dwarf, which are known to have large flares with a duty cycle of $\sim 1$\%  \citep{2007AJ....133.2258S}.

\section{Conclusions}

We have identified two nearby proper motion stars at low galactic latitudes, demonstrating the value of an infrared proper motion survey based on the WISE and 2MASS catalogs. \citet{2010AJ....140.1868W} note that an important WISE mission objective is to identify unknown, nearby low-luminosity brown dwarfs. Proper motion selection using WISE promises to also contribute to the solar neighborhood {\it stellar} census over the entire sky, including the Galactic Plane. The objects detected here are consistent with the evidence \citep{2005AJ....129.1483L,2008AJ....135.2177L} that the vast majority of high proper motion hydrogen-burning stars have been detected. \citet{2005AJ....130.1680L}, however, estimates that a third of nuclear-burning stars within 33 parsecs remain unidentified but have motions below 0.15 "/yr. WISE and 2MASS photometry and astrometry appear to be accurate enough to allow many of these to be identified, even at low Galactic latitudes, although optical follow-up will be required.

\acknowledgments

We thank the Annie Jump Cannon Fund at the University of Delaware for support and the anonymous referee for useful comments.

This publication makes use of data products from the Wide-field Infrared Survey Explorer, which is a joint project of the University of California, Los Angeles, and the Jet Propulsion Laboratory/California Institute of Technology, funded by the National Aeronautics and Space Administration. This publication makes use of data products from the Two Micron All Sky Survey, which is a joint project of the University of Massachusetts and the Infrared Processing and Analysis Center/California Institute of Technology, funded by the National Aeronautics and Space Administration and the National Science Foundation. This research has made use of the NASA/ IPAC Infrared Science Archive, which is operated by the Jet Propulsion Laboratory, California Institute of Technology, under contract with NASA. This research has made use of the VizieR catalogue access tool, CDS, Strasbourg, France. This research has made use of the SIMBAD database, operated at CDS, Strasbourg, France. The Digitized Sky Surveys were produced at the Space Telescope Science Institute under U.S. Government grant NAG W-2166. The images of these surveys are based on photographic data obtained using the Oschin Schmidt Telescope on Palomar Mountain and the UK Schmidt Telescope. The plates were processed into the present compressed digital form with the permission of these institutions.

%I don't think this is necessary but here it is for those who look at comments:

%The National Geographic Society - Palomar Observatory Sky Atlas (POSS-I) was made by the California Institute of Technology with grants from the National Geographic Society.

%The Second Palomar Observatory Sky Survey (POSS-II) was made by the California Institute of Technology with funds from the National Science Foundation, the National Geographic Society, the Sloan Foundation, the Samuel Oschin Foundation, and the Eastman Kodak Corporation.

%The Oschin Schmidt Telescope is operated by the California Institute of Technology and Palomar Observatory.

The UK Schmidt Telescope was operated by the Royal Observatory Edinburgh, with funding from the UK Science and Engineering Research Council (later the UK Particle Physics and Astronomy Research Council), until 1988 June, and thereafter by the Anglo-Australian Observatory. The blue plates of the southern Sky Atlas and its Equatorial Extension (together known as the SERC-J), as well as the Equatorial Red (ER), and the Second Epoch [red] Survey (SES) were all taken with the UK Schmidt.

%tab
\begin{deluxetable}{lrlrrrrrrrr}
\tablewidth{0pc}
\tabletypesize{\footnotesize}
\tablenum{1}
\tablecaption{2MASS and WISE Data}
\tablehead{
\colhead{WISEP Name} & \colhead{$\mu_{\alpha}$} &
\colhead{$\mu_{\delta}$  ("/yr)} &
\colhead{J} &
\colhead{H} &
\colhead{K$_s$} &
\colhead{W$_1$} & 
\colhead{W$_2$} & 
\colhead{W$_3$} &
\colhead{W$_4$}  }
\startdata
J191239.91$-$361516.4 & 0.78 & -1.94 & 9.52 & 9.01 & 8.77 & 8.55 & 8.35 & 8.20 & 8.17 \\
J190648.47+401106.8 & 0.44 & -0.18 & 13.08 & 12.26 & 11.77 & 11.45 & 11.22 & 10.77 & $>9.32$ \\
\enddata
\end{deluxetable}

\end{document}